\documentclass[universe,review,accept,pdftex,moreauthors]{Definitions/mdpi} 
\firstpage{1} 
\pubvolume{8}
\issuenum{10}
\articlenumber{493}
\pubyear{2022}
\copyrightyear{2022}
\externaleditor{{Academic Editor: Vikram Dwarkadas} %MDPI: please add.
}
\datereceived{07 July 2022} 
\dateaccepted{14 September 2022} 
\datepublished{21 September 2022} 
\hreflink{https://doi.org/ } % If needed use \linebreak
\usepackage{graphicx}
\usepackage{subfig}
\newcommand{\msun}{\mbox{M$_{\odot}$}}

\newcommand{\kms}{\mbox{$\rm{km}\,s^{-1}$}\,}

\DeclareMathAlphabet{\mathsc}{OT1}{cmr}{m}{sc}
\def\testbx{bx}%
\DeclareRobustCommand{\ion}[2]{%
\relax\ifmmode
\ifx\testbx\f@series
{\mathbf{#1\,\mathsc{#2}}}\else
{\it{#1\,\mathsc{#2}}}\fi
\else\textup{#1\,{\mdseries\textsc{#2}}}%
\fi}
\newcommand{\Ha} {\mbox{H$\alpha$}\,}

\newcommand{\apjs} {ApJS,}

\Title{Gap Transients Interacting with Circumstellar Medium}

\TitleCitation{Gap Transients Interacting with Circumstellar Medium}

\Author{Yongzhi Cai %MDPI: Please carefully check the accuracy of names and affiliations.
 $^{1,}$*\orcidA{}, Andrea Reguitti $^{2,3,4}$\orcidB{}, Giorgio Valerin $^{4,5}$\orcidC{} and Xiaofeng Wang $^{1,6,}$*}

\AuthorNames{Yongzhi Cai, Andrea Reguitti, Giorgio Valerin and Xiaofeng Wang}

\AuthorCitation{Cai, Y.; Reguitti, A.; Valerin, G.; Wang, X.}
\address{%
$^{1}$ \quad \textls[-15]{Physics Department and Tsinghua Center for Astrophysics (THCA),
Tsinghua University, \mbox{Beijing 100084, China}}\\
$^{2}$ \quad Departamento de Ciencias Fisicas, Universidad Andres Bello, Fernandez Concha 700, Las Condes, Santiago~8320000, %MDPI: Please add the postal code (or ZIP code in the U.S.). If the postal code is not available, please provide the P.O. Box.
Chile\\
$^{3}$ \quad Millennium Institute of Astrophysics (MAS), Nuncio Monsenor S\`{o}tero Sanz 100, Providencia, Santiago~8320000, Chile  \\
$^{4}$ \quad INAF---Osservatorio Astronomico di Padova, Vicolo dell’Osservatorio 5, 35122 Padova, Italy\\
$^{5}$ \quad Dipartimento di Fisica e Astronomia, %MDPI: we change the  address information from subordinate to superior, please confirm.
 Universit\`a degli Studi di Padova, Vicolo dell'Osservatorio 2, 35122~Padova, Italy \\
$^{6}$ \quad Beijing Planetarium, Beijing Academy of Science and Technology, Beijing, 100044, China\\
}
\corres{Correspondence: yzcai789@163.com (Y.C.); wang\_xf@mail.tsinghua.edu.cn (X.W.)}

\abstract{In the last 20 years, modern wide-field surveys discovered a new class of peculiar transients, which lie in the luminosity gap between standard supernovae and classical novae. These transients are often called “intermediate luminosity optical transients” %MDPI: if we should keep not only the italic, but also the font of the word?
or “gap transients”. They are usually distinguished in subgroups based on their phenomenology, such as supernova impostors, intermediate luminosity red transients, and luminous red novae. In this review, we present a brief overview of their observational features and possible physical scenarios to date, in the attempt to understand their nature.
}

\keyword{supernovae; diversity; progenitor; explosion mechanism} % (List three to ten pertinent keywords specific to the article; yet reasonably common within the subject discipline.)

\begin{document}

%%%%%%%%%%%%%%%%%%%%%%%%%%%%%%%%%%%%%%%%%%
%\setcounter{section}{-1} %% Remove this when starting to work on the template.
%\section{How to Use this Template}

\section{Introduction}
 Beyond %MDPI: we remove the bold, please confirm.
 supernovae (SNe), few cosmic events can release an amount of kinetic energy of the order of 10$^{51}$ erg (1 foe). On the higher end of the energy spectrum, we find Superluminous SNe (SL SNe; e.g., %MDPI: please confirm the changed format.
 \cite{Inserra2019NatAs...3..697I,GalYam2019ARA&A..57..305G,Chen2022arXiv220202059C,Chen2022arXiv220202060C}), which are extreme examples of stellar explosions, and in some cases, Fast Blue Optical Transients (FBOTs; e.g., \cite{Drout2014ApJ...794...23D, Margutti2019ApJ...872...18M, Ho2021arXiv210508811H,Fraser2021arXiv210807278F,Xiang2021ApJ...910...42X,Metzger2022ApJ...932...84M}) can even exceed the energy released by SL SNe. On the other hand, a growing number of stellar transients release less kinetic energy than the typical SNe, while still being more luminous than classical Novae. The names {\sl {Intermediate luminosity optical transients}} \cite{Berger2009ApJ...699.1850B, Soker2012ApJ...746..100S, Soker2021RAA....21...90S} or {\sl {gap transients}} refer to such underluminous events. 
The early discoveries of {\sl {gap transients}} were made a few decades ago (e.g., M31-RV, SN 1997bs; see \cite{VanDyk2000PASP..112.1532V, Boschi2004A&A...418..869B}); today, after the introduction of wide-field synoptic surveys, such as the Zwicky Transient Facility (ZTF; e.g., \cite{Graham2019PASP..131g8001G}), the Asteroid Terrestrial-impact Last Alert System (ATLAS; e.g., \cite{Tonry2018PASP..130f4505T}), and the Panoramic Survey Telescope and Rapid Response System (Pan-STARRS; e.g., \cite{Chambers2016arXiv161205560C}), the number of transients populating this luminosity gap keeps growing: it is time to provide a clearer picture of these phenomena.

\begin{figure}[H]
\includegraphics[width=13cm]{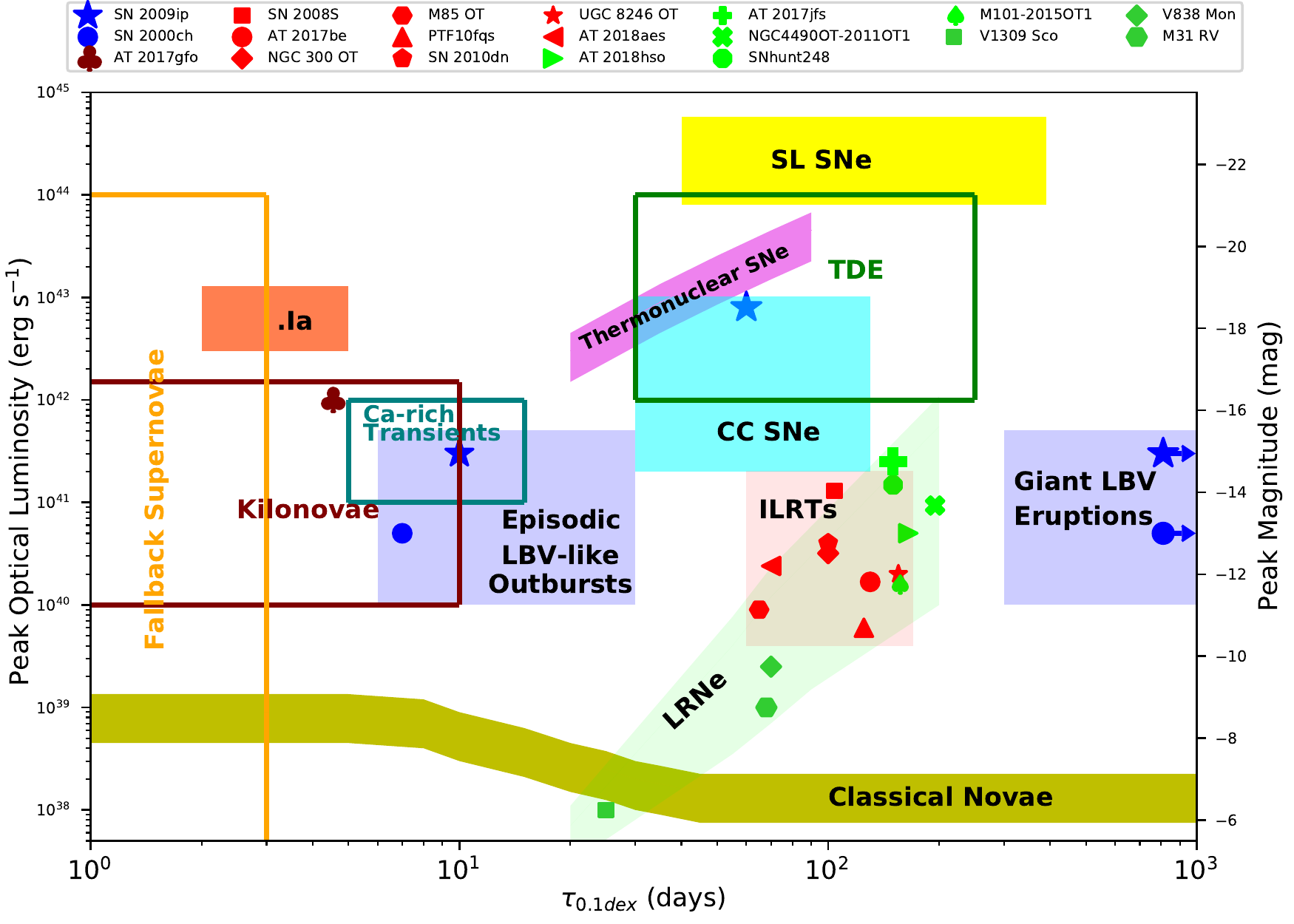}
\caption{Peak {optical} luminosity vs. characteristic time (defined as the time for a luminosity decline of 0.1 dex) for optical transients in the local Universe. {Various regions, marked by different colours and shapes, indicate the general location of some representative classes of transients (e.g., superluminous SNe (SL SNe), tidal disruption event (TDE), Thermonuclear and core-collapse SNe (CC SNe), classical novae, Ca-rich transients, SNe .Ia, Kilonovae, fallback SNe, events from LBVs, ILRTs, LRNe).} Figure adapted from Rau et al. \cite{Rau2009PASP..121.1334R}, Kasliwal et al. \cite{Kasliwal2011ApJ...730..134K}. \label{fig1}}
\end{figure}

{In the poorly known luminosity range known as the “gap”} \cite{Pastorello2019NatAs...3..676P}, we find several types of stellar transients with heterogeneous observational properties and various physical {origins}. This luminosity gap contains the giant eruptions of massive stars including luminous blue variables (LBVs), intermediate-luminosity red transients (ILRTs), and luminous red novae (LRNe) {(see Figure~\ref{fig1})}.
Although these gap transients belong to different subtypes, {their observational features are often similar}; hence, their classification can be challenging. {They also share physical characteristics, such as being on the Optical Transient Stripe (see e.g., \cite{Kashi2010arXiv1011.1222K, Kashi2016RAA....16...99K, Kashi2017MNRAS.468.4938K, Soker2016MNRAS.462..217S}).} This motivated us to systematically summarise their observations and physics to date. 

% Example of a figure that spans the whole page width. The same concept works for tables, too.

%%%%%%%%%%%%%%%%%%%%%%%%%%%%%%%%%%%%%%%%%%
\section{Supernova Impostors}

Massive ($\geq$40 $M_{\odot}$) stars close to the end of their evolutionary paths can experience luminous, non-terminal eruptions, {in which they can lose from tenths to tens of solar masses~\cite{Smith2006ApJ...645L..45S}}. Because their spectra and photometric properties mimic those of genuine Type-IIn SNe \cite{Schlegel1990MNRAS.244..269S,Filippenko1997ARA&A..35..309F}, but the progenitors survived the events, these gap transients are often dubbed as “Supernova impostors” \cite{VanDyk2000PASP..112.1532V,Maund2006MNRAS.369..390M}.
These events are usually associated with Luminous Blue Variables (LBVs; \cite{Humphreys1994PASP..106.1025H}).
LBVs are a short phase in the final evolution of evolved and massive stars {\cite{Maeder1994ARA&A..32..227M}}, during which they can reach and exceed a luminosity of 10$^5$ $L_{\odot}$.
{In recent times a lot of effort has been devoted to solve the disagreement} {regarding the ‘sociality’ of LBVs, i.e., if they tend to be isolated stars or confined in binary systems, and whether they are associated or not with other young massive stars (see, e.g., \cite{Smith2015MNRAS.447..598S, Aghakhanloo2017MNRAS.472..591A, Humphreys2016ApJ...825...64H, Aadland2018AJ....156..294A}, for discording results). This discrepancy has profound implications on their evolution, both regarding their mass loss history and age. } 
%Nonetheless we have numerous examples of LBVs that are confirmed multiple systems, such as the already mentioned Eta Car, or HD 5980, which is even a triple system with a WR and an O-type star as companions (Koenigsberger et al. 2010).

\begin{figure}[H]
\begin{adjustwidth}{-\extralength}{0cm}
\centering %% If there is a figure in wide page, please release command \centering
\includegraphics[width=9cm]{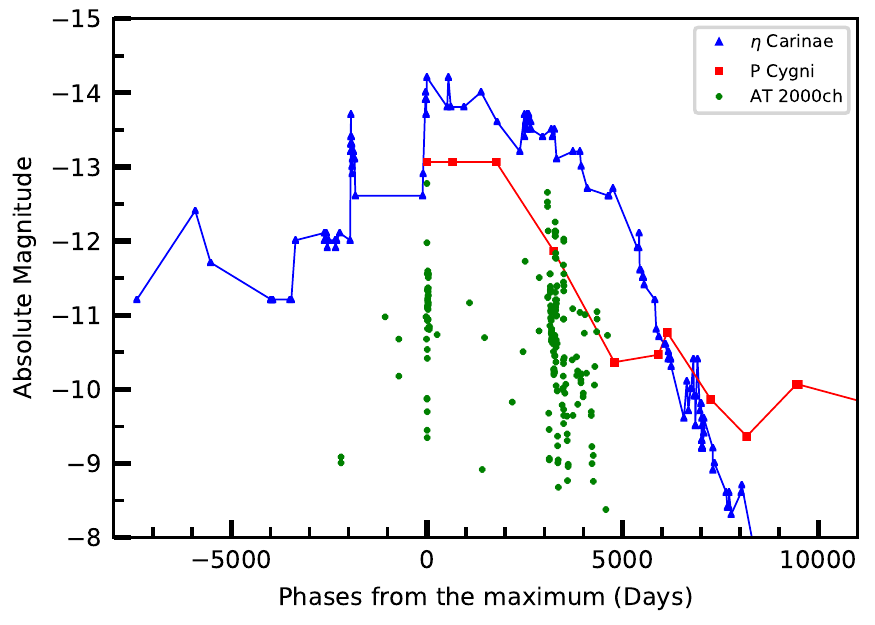}
\includegraphics[width=9cm]{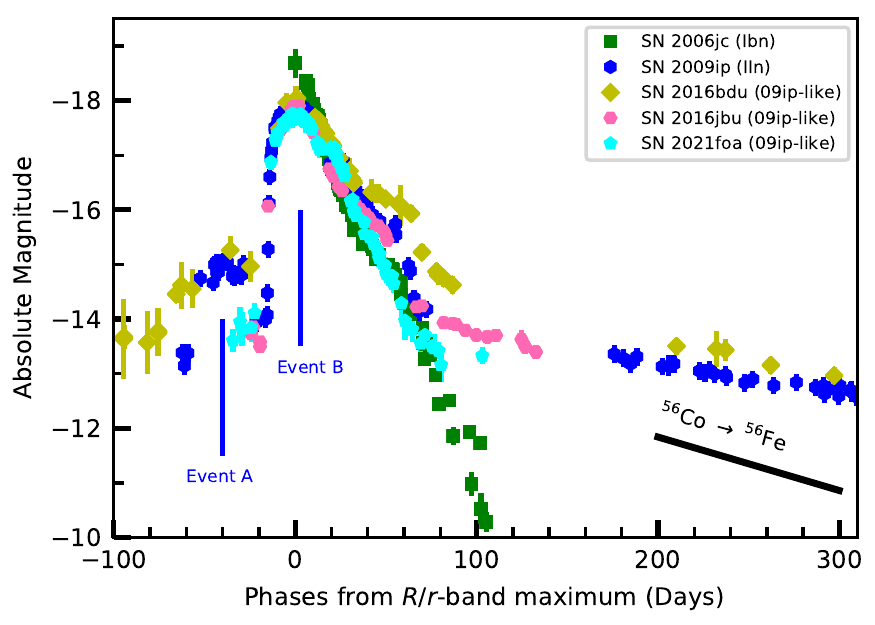}
\end{adjustwidth}
\caption{\textls[-15]{\textbf{Left %MDPI: we change the subfigure format to left and right, please confirm?
%MDPI: Please use commas to separate thousands for numbers with five or more digits (not for four digits) in the picture. e.g., "10000" should be "10,000"
} panel: $V/R$-band absolute light curves of {historical} LBV eruptions. \textbf{Right %MDPI: please confirm if this should be botttom?
} panel: $R/r$-band absolute light curves of SN2009ip-like objects with pre-supernova outbursts. 
{The light curve peaks of SN 2009ip, indicated as ‘Event A’ and ‘B’, %MDPI: we add the symbold here, please confirm.
 are marked.
The expected decline rate from the radioactive decay of $^{56}$Co into $^{56}$Fe is also reported. The late-time decline rate of the light curves is shallower than the one expected from simple $^{56}$Co decay, therefore suggesting the presence of an additional powering source in the form of CSM interaction.}
Data from \citet{Frew2004JAD....10....6F, Foley2007ApJ...657L.105F,Smith&Frew2011MNRAS.415.2009S,Smith2011MNRAS.415..773S}, Pastorello et al. \cite{Pastorello2007Natur.447..829P_06jc, Pastorello2010MNRAS.408..181P,Pastorello2013ApJ...767....1P, Pastorello2018MNRAS.474..197P}, \citet{Fraser2013MNRAS.433.1312F,Fraser2015MNRAS.453.3886F, Brennan2022MNRAS.513.5642B,Brennan2022MNRAS.513.5666B, Reguitti2022A&A...662L..10R}.}
\label{SNimpostors} }
\end{figure}

LBVs can experience bright and years-long “giant eruptions”, the most famous example being the luminous event from the Galactic LBV $\eta$ Carinae (see the left panel of Figure~\ref{SNimpostors}) in the middle of 19th century \cite{Humphreys1994PASP..106.1025H, Soker2001A&A...377..672S, Soker2001MNRAS.325..584S}, during which the star reached an impressive absolute magnitude of $-$14 mag \cite{Smith&Frew2011MNRAS.415.2009S,Davidson2012Natur.486E...1D}, shortly becoming  the second brightest star in the night sky and releasing an amount of energy comparable to that of an SN event~\cite{Smith2006ApJ...645L..45S}; however, it survived the event.
{Smith \cite{Smith2011MNRAS.415.2020S} proposed a model in which a violent collision of stars in an eccentric binary system at the periastron may have been the cause of the Great Eruption of $\eta$ Carinae around the 1840s.}
Another Galactic star recognised as an LBV is P Cygni (in which the homonym spectral line profile was {firstly observed}), which experienced repeated outbursts around four centuries ago \cite{DeGroot1988IrAJ...18..163D}. {Interestingly, P Cygni was also suggested to be a binary system \cite{Kashi2010MNRAS.405.1924K}. }
%61V woosley 2022
During the great eruptions, the star's mass loss rate increases significantly, becoming greater than 10$^{-4}$/10$^{-3}$ \msun yr%MDPI: please confirm if \msun should be italic and yr should be year, please check for the whole text.
$^{-1}$ \cite{Smith2006ApJ...638.1045S}, even of the order of 10$^{-1}$ \msun yr$^{-1}$ for $\eta$ Carinae, and several to tens of solar masses are expelled during its entire duration ({\cite{Smith2003AJ....125.1458S}, from the mass of the Homunculus Nebula, created after the 1843 Event).}

%%%% Light curves of SN Impostors

Occasionally, extragalactic giant eruptions can be observed in external nearby galaxies, such as {AT 2000ch} \citep{Wagner2004PASP..116..326W}, a well-known LBV in the galaxy NGC 3432, which generated multiple bright outbursts (up to $M_R$$\sim$$-14$ mag) since its discovery (see the left panel of Figure~\ref{SNimpostors}), most notably in 2008 and 2009 \citep{Pastorello2010MNRAS.408..181P}.
Other relevant examples include the SN 1954J \citep{Tammann1968ApJ...151..825T} and SN 2002kg \citep{Weis2005A&A...429L..13W} events, identified as major brightenings of two known LBVs in the galaxy NGC 2403 {\citep{VanDyk2005PASP..117..553V,Humphreys2017ApJ...848...86H}}, and SN 1997bs \citep{VanDyk2000PASP..112.1532V}, an analogue of the $\eta$ Carinae  eruption in M66.
In some occasions, isolated outbursts from massive hypergiants/LBVs are also observed \citep{Tartaglia2015MNRAS.447..117T,Tartaglia2016ApJ...823L..23T}.

%The high csm density cannot be explained by standard wind mass-loss lamers 95,98

\begin{figure}[H]
\includegraphics[width=13.5cm]{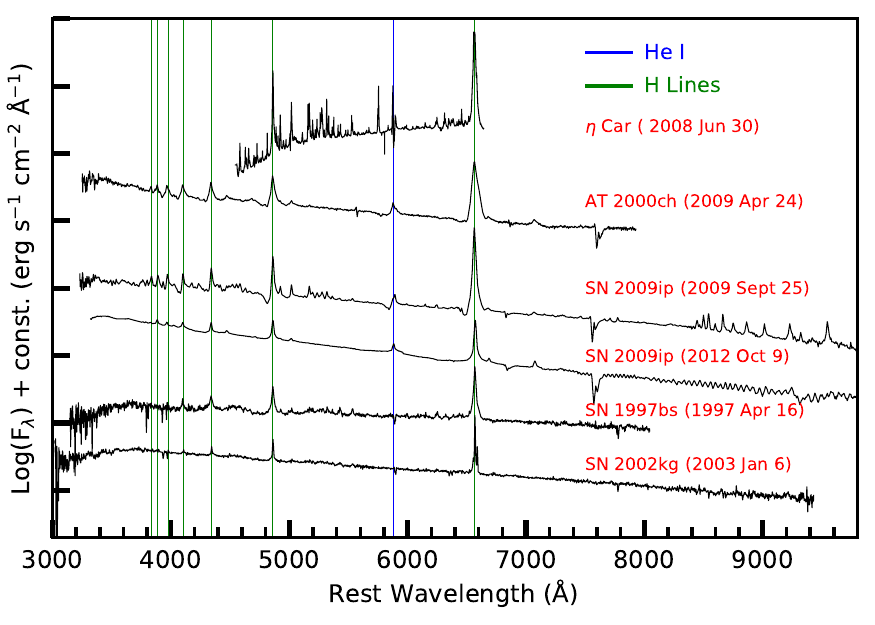}
\caption{{Spectra of LBV eruptions and SN impostors} at different phases. {All the spectra show a blue continuum and narrow H emission lines, often with P Cygni profiles. Sometimes, He I lines are also identified. The principal lines from H and He are marked with vertical lines.} \label{figSNImspectra}}
\end{figure}

The spectra of such events reveal narrow ($\sim$10$^2$--10$^3$ \kms, %MDPI: please confirm if \cdot should be added between the two units.
 {\citep{Smith2017hsn..book..403S}}) emission lines {from the Balmer series} on top of blue continua {(see some important examples in Figure \ref{figSNImspectra})}.
The narrow emission lines result from the photoionization of the {unshocked} circumstellar medium {(CSM)} surrounding the star {\citep{Chugai1994MNRAS.268..173C}}.
It is still unclear how  major mass-loss episodes can be triggered.
Stellar encounters in close binary systems is a viable explanation for some Giant Eruptions, as $\eta$ Carinae itself is in a binary system \citep{Damineli1996ApJ...460L..49D}.
For single stars, instead, extreme super-Eddington continuum-driven winds have also been invoked \citep{Owocki2004ApJ...616..525O}, although it is unknown what can lead to a sudden and conspicuous excess in the star's Eddington luminosity/mass limit {(see also \cite{Lamers1988ApJ...324..279L})}. {One possibility is a variation in opacity at the helium peak beneath the surface \citep{Jiang2018Natur.561..498J}.}

\vspace{-6pt}
%%%% Spectra of SN Impostors

\subsection*{Pre-Supernova Outbursts}
In an increasing number of cases, one or more SN impostors were observed a few months to years before the explosion of a real SN (e.g., \cite{Ofek2014ApJ...789..104O,Reguitti2019MNRAS.482.2750R,Strotjohann2021ApJ...907...99S} and references therein). Most of these SNe belong to the class of type IIn \citep{Schlegel1990MNRAS.244..269S,Filippenko1997ARA&A..35..309F}, whose spectra are dominated by a blue continuum, superimposed by narrow (FWHM of some 10$^2$ \kms, up to a few 10$^3$~\kms) emission lines of H, features are also common in Supernova Impostors spectra.
These lines are indicative of interaction of fast SN ejecta with slow-moving H-rich CSM ejected shortly before {(tens to thousands of years \citep{Smith2014ARA&A..52..487S})} the SN explosion {(\citep{Chevalier1994ApJ...420..268C}, for a review of the  ejecta-CSM interaction theory)}.\\

\textls[-15]{The most important object with multiple outbursts preceding a SN IIn is SN 2009ip~\citep{Smith2010AJ....139.1451S,Margutti2014ApJ...780...21M}.} It was discovered on August 2009 at an absolute magnitude of $M_V\approx-15$ mag, but despite the name, the 2009 event was not a terminal one.
The progenitor was recovered months after, and other luminous episodes were seen in 2011 (see the right panel of \mbox{Figure~\ref{SNimpostors}}). On July 2012, another outburst reaching the same luminosity level as the 2009 one ($M_V$$\sim$$-15$) was observed. Finally, on October 2012, the brightest and much more luminous brightening occurred, with an absolute $-$18 mag. Then, its luminosity has been declining ever since, and no other major outbursts have been detected.
In fact, nowadays {SN 2009ip} is even fainter than the quiescent progenitor \citep{Smith2022arXiv220502896S} identified by \citet{Foley2011ApJ...732...32F} at an absolute magnitude of $-$10 (consistent with a star of 50--60 \msun %MDPI: please confirm if this should be italic?
) in archival HST (Hubble Space Telescope) %MDPI: Please confirm if the italics should be retained.
 images taken years before the 2009 event. {\citet{Soker2013ApJ...764L...6S} suggested that the progenitor of SN 2009ip was a massive binary system where the primary star was {a} massive LBV of \mbox{60--100 {\msun}} with a much lighter main-sequence star as a companion.}

{There has been some debate regarding} the nature of the two events {of SN 2009ip} observed in 2012, using the terminology in \citet{Pastorello2013ApJ...767....1P}; they are called ‘Event~A’ and ‘Event B’, respectively.
\citet{Mauerhan2013MNRAS.431.2599M} and \citet{Graham2014ApJ...787..163G} {argue that ‘Event~A’} was a true SN event, and the strong interaction of the fast ejecta that collided with a pre-existing shell of CSM material ejected during the previous LBV-like eruptive episodes powered ‘Event B’. This scenario has been recently modelled by \citet{Chugai2022arXiv220905204C}. 
However, \citet{Fraser2013MNRAS.433.1312F} and \citet{Pastorello2013ApJ...767....1P} disagree on that: while ejecta velocities of 10$^4$~\kms are normal in CCSNe, they also observed material moving at 13,000~\kms during the 2011 event, which was surely  \textit{not} the end of the progenitor star. In addition, the ‘Event A’ luminosity was exceptionally dim for a true SN (but see \citep{Woosley2007Natur.450..390W}), and no newly synthesized material was seen.
Instead, they suggest that the ‘Event A’ was powered by the Pulsational Pair Instability mechanism \citep{Woosley2007Natur.450..390W,Woosley2017ApJ...836..244W}, followed by collision with pre-existing CSM sustaining the ‘Event B’, and the star may ultimately have survived.
%This scenario is more likely given the faint abs mag of the first Event, while the second had a luminosity more consistent with a true CCSN.
{Finally, another model proposed by \citet{Kashi2013MNRAS.436.2484K} for the outbursts of SN 2009ip is a repeated binary interaction at periastron passages, with the two stars finally merging during the second 2012 outburst \citep{Soker2013ApJ...764L...6S}, in a similar fashion as the \citet{Smith2011MNRAS.415.2020S} model for $\eta$~Carinae.}

In the years since, other objects with photometric and spectroscopic similarities to SN 2009ip (dubbed as 'SN 2009ip-like' objects) have been discovered \citep{Pessi2022ApJ...928..138P,EliasRosa2016MNRAS.463.3894E,Thone2017A&A...599A.129T,Pastorello2018MNRAS.474..197P,Tartaglia2016MNRAS.459.1039T,Brennan2022MNRAS.513.5642B,Brennan2022MNRAS.513.5666B}, and recently even a transitional object between SN\, 2009ip-like and Ibn SNe \citep{Reguitti2022A&A...662L..10R}, but the relatively high rate of these objects has raised the question if all those transients are indeed produced by massive stars \citep{Brennan2022MNRAS.513.5666B}.
Recently, very late-time $HST$ observations of SN 2009ip-like events SN 2015bh \citep{Jencson2022arXiv220602816J} and SN 2016jbu \citep{Brennan2022arXiv220606365B} revealed that the transients are now much fainter than their progenitors, supporting the idea that SN 2009ip-like objects are indeed genuine, albeit strange, {terminal} SNe.

\begin{figure}[H]
\centering
\includegraphics[width=12cm]{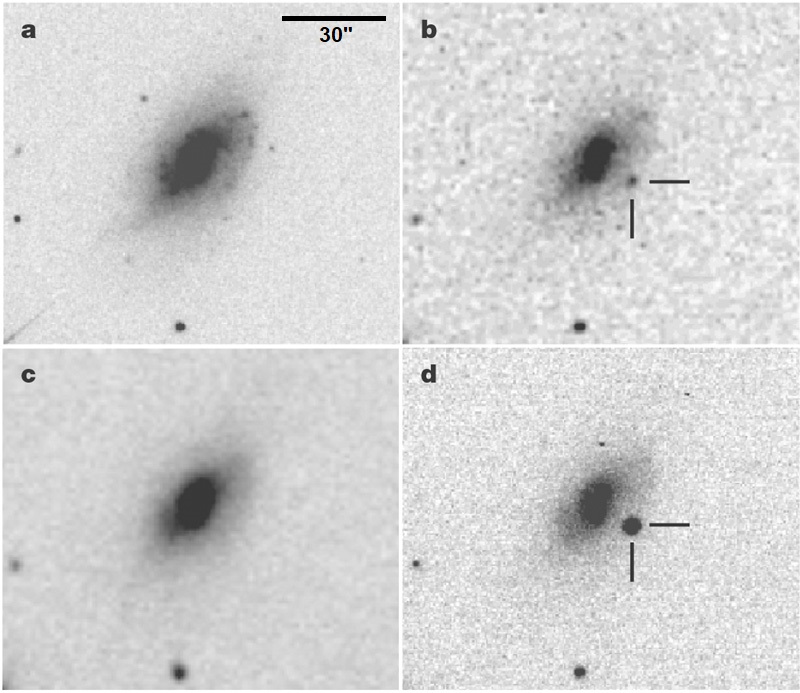}
\caption{(\textbf{a}) A Sloan Digital Sky Survey image of UGC 4904 (host galaxy of SN 2006jc) obtained in 2001. No transient is visible. (\textbf{b}) The faint transient UGC 4904-V1 on 16 October 2004. (\textbf{c}) The same field on 21 September 2006. No transient is visible. (\textbf{d}) SN 2006jc on 2006 October 29. The position of UGC 4904-V1 is coincident with that of SN 2006jc, making it an SN impostor prior to the real SN event. Figure from \citet{Pastorello2007Natur.447..829P_06jc}. \label{fig3}}
\end{figure}

Nonetheless, SN impostors are not only observed before a type-IIn SN or from LBVs. For instance, in at least one case, a pre-SN outburst was spotted before a type-Ibn SN. 
A $M_R$$\sim$$-14$ mag outburst was observed 2 years prior to the appearance of the type-Ibn SN 2006jc \citep{Foley2007ApJ...657L.105F,Pastorello2007Natur.447..829P_06jc}, see Figure \ref{fig3}.
Differently from type-IIn SNe, SNe Ibn present no H lines, but narrow He emission lines instead \citep{Pastorello2008MNRAS.389..131P}.
The precursor event was likely produced by a Wolf--Rayet star during an episode of intense mass-loss, and it created a dense He-rich CSM around the progenitor, {with} which  the ejecta from the SN later interacted.

Bright outbursts immediately before an SN explosion are not predicted by current models of stellar evolution, as late evolutionary stages are difficult to simulate.
Nonetheless, a few physical mechanisms have been proposed to explain pre-SN eruptions, such as violent convection and unstable late nuclear
burnings \citep{Arnett2011ApJ...741...33A,Quataert2012MNRAS.423L..92Q,Shiode2014ApJ...780...96S}.

With the study of pre-SN outbursts and the comprehension of which processes are their drivers, it will be possible to establish if these events can herald the imminent core-collapse of massive stars.

%%%%%%%%%%%%%%%%%%%%%%%%%%%%%%%%%%%%%%%%%%
\section{Intermediate Luminosity Red Transients}
\begin{figure}[H]
%\begin{adjustwidth}{-\extralength}{0cm}
\includegraphics[width=11.6cm]{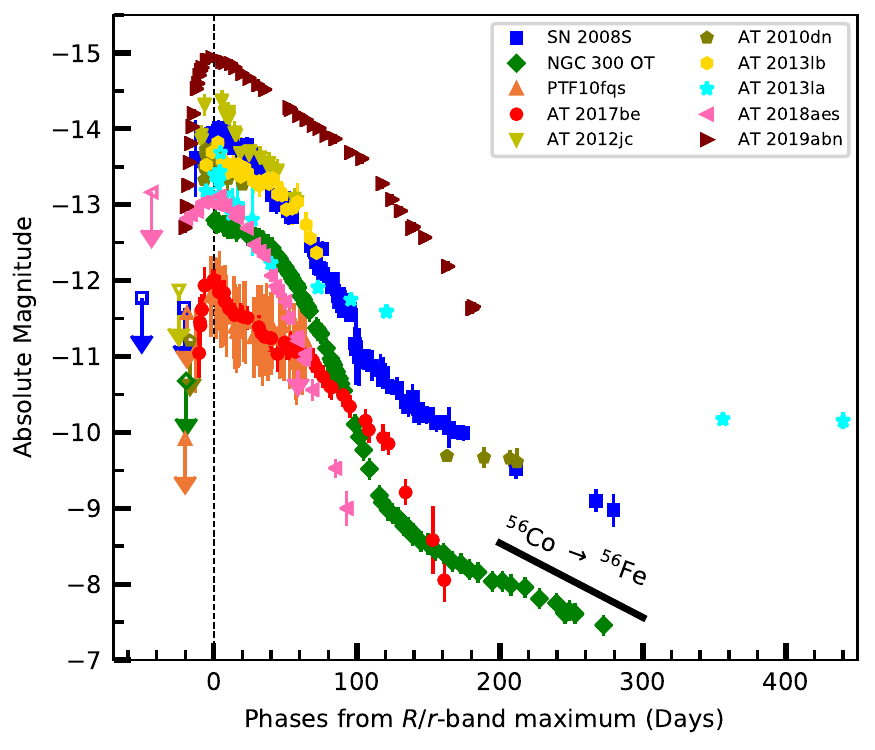}\par
\includegraphics[width=12cm]{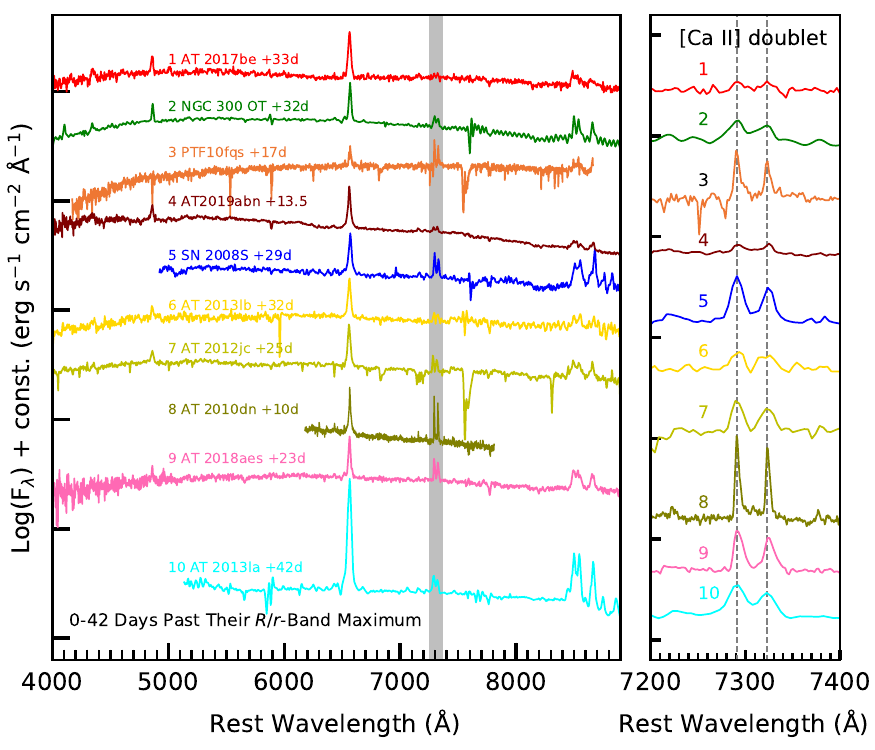}\par
%\end{adjustwidth}
\caption{\textbf{Top} panel: $R/r$-band absolute light curves of ILRTs. \textbf{Bottom} panel: spectral comparison of ILRTs, along with the [Ca II] region {which is marked in grey}. {Their [Ca II] lines ($\lambda\lambda$ 7291, 7324) are enlarged on the right of this panel.} 
Data from \citet{Botticella2009MNRAS.398.1041B, Bond2009ApJ...695L.154B,Kasliwal2011ApJ...730..134K,Humphreys2011ApJ...743..118H, Cai2018MNRAS.480.3424C,Cai2021A&A...654A.157C,Williams2020A&A...637A..20W,Stritzinger2020A&A...639A.103S}.
\label{ILRTs}}
\end{figure}

Intermediate Luminosity Red Transients (ILRTs) are a class of poorly studied gap transients that sparked debate over their origin and interpretation. 
While some authors invoked non-terminal outbursts of post main sequence stars to explain the observed data (e.g., \citep{Kashi2010ApJ...709L..11K,Humphreys2011ApJ...743..118H}), ILRTs are among the most promising candidates for being Electron Capture Supernova (EC SN) events {(see, e.g., \citep{Botticella2009MNRAS.398.1041B,Cai2018MNRAS.480.3424C,Pastorello2019NatAs...3..676P, Cai2021A&A...654A.157C})}.
Their light curves are reminiscent of classical SNe (see the left panel of Figure~\ref{ILRTs}), with a single peak and a late-time linear decline in magnitudes compatible with $^{56}$Ni decay \citep{Botticella2009MNRAS.398.1041B,Cai2018MNRAS.480.3424C,Cai2021A&A...654A.157C}. Indeed, long-term mid-infrared monitoring displayed that two ILRTs (NGC300 2008OT-1 and SN 2008S) became fainter than their progenitors years after their peak luminosity, corroborating the scenario of a terminal explosion rather than an eruptive event \citep{Adams2016MNRAS.460.1645A}.

Despite the indicators that ILRTs are indeed the result of a stellar explosion, differentiating between a faint Fe CC event \citep{Pastorello2004MNRAS.347...74P,Spiro2014MNRAS.439.2873S,Reguitti2021MNRAS.501.1059R,Valerin2022MNRAS.513.4983V} and an EC SN \citep{Nomoto1984ApJ...277..791N,Poelarends2008ApJ...675..614P,Pumo2009ApJ...705L.138P,Moriya2014A&A...569A..57M,Doherty2015MNRAS.446.2599D} is not an easy task.
First of all, EC SNe are expected to be weak explosions following the collapse of an O--Ne--Mg core, therefore yielding low peak luminosity as well as slowly expanding ejecta, compared to standard SN events \citep{Janka2008A&A...485..199J}. Lying within the luminosity gap, ILRTs easily fulfill the condition of being low-energy explosions.
Additionally, EC SNe should synthesize only few 10$^{-3}$ M$_{\odot}$ of $^{56}$Ni \citep{Wanajo2009ApJ...695..208W}, which, again, is consistent with observational constraints obtained for ILRTs \citep{Botticella2009MNRAS.398.1041B,Cai2021A&A...654A.157C}.
Finally, the progenitor star of an EC SN is expected to be a \mbox{$\sim$9 M$_{\odot}$} luminous super-Asymptotic giant branch (AGB) star, which, during its evolution, will develop a degenerate O--Ne--Mg core \citep{Poelarends2008ApJ...675..614P}. This is a key condition, but at the same time, it is the most difficult to probe, given that only for targets closer than $\sim$10 Mpc is it possible to study the progenitor star in archival images. In the few cases of ILRT with a successful detection in archival images, the results are encouraging: the progenitor stars of NGC 300 2008OT-1 and SN 2008S were identified as extreme AGB stars, enshrouded in dust and with masses between 8 and 12 M$_{\odot}$ \cite{Thompson2009ApJ...705.1364T}. While such detailed identification was impossible for the progenitor of AT 2019abn, the results and upper limits obtained were still compatible with a super-AGB star \citep{Jencson2019ApJ...880L..20J}.
It is also worth noticing that the estimated rate of ILRTs is $\sim$8\% of the total core collapse events, perfectly in line with what is expected for EC SNe \citep{Cai2021A&A...654A.157C}. 

To better understand the observational properties of these transients, it is useful to discuss the case of the closest ILRT ever studied, NGC 300 2008OT-1, which is located at less than 2 Mpc from the Milky Way and proves to be a valuable benchmark for the whole class of objects \citep{Bond2009ApJ...695L.154B,Berger2009ApJ...699.1850B,Humphreys2011ApJ...743..118H}.
The light curve of NGC 300 2008OT-1 is missing the rising phase due to solar conjunction, but its overall shape is similar to the light curves of SNe IIP, with a slow decline in the first phases followed by a sharper drop in luminosity, and finally, a linear decline compatible with $^{56}$Ni decay. The spectra are dominated by narrow Balmer emission lines, with a full width at half maximum of $\sim$1000 km s$^{-1}$. Ca lines are also prominent spectral features, both in absorption, such as Ca H\&K at early phases, and in emission, such as the forbidden [Ca II] doublet and Ca Near-Infrared (NIR) triplet. In particular, the [Ca {\sc ii}] doublet ($\lambda\lambda$ 7291, 7324) feature, which is visible at all phases, is a characteristic signature for ILRTs%MDPI: we change footnote to Note format, please confirm.
\endnote{{However, weak [Ca II] lines were  also detected in the spectra of the LRN AT 2018hso \citep{Cai2019A&A...632L...6C}.}} (see the right panel of Figure~\ref{ILRTs}).
The study of the spectral continuum over time reveals that the emitting source is monotonously cooling after maximum luminosity. NIR observations suggest the presence of thermal emission from dust grains, which formed once the temperature in the expanding gas became sufficiently low to allow condensation~\citep{Prieto2009ApJ...705.1425P,Ohsawa2010ApJ}. An additional dust component, cooler and more distant from the transient, has been identified~\citep{Humphreys2011ApJ...743..118H}, possibly pre-existing dust that survived the explosion. The geometry of dust has been discussed in detail for SN 2008S, which displayed similar observables, but more generally, dust formation appears to be a ubiquitous characteristic in ILRTs \citep{Botticella2009MNRAS.398.1041B,Cai2021A&A...654A.157C}.
Thanks to its proximity, NGC 300 2008OT-1 was also studied through high-resolution spectroscopy, revealing narrow ($\sim$40 km s$^{-1}$) absorption lines superimposed to the H$\alpha$ line and Ca NIR triplet, suggesting the presence of slowly moving gas shells around the transient \citep{Berger2009ApJ...699.1850B}. Other identified lines include Fe II and He I, as well as narrow O I and Na I.
The complete absence of a P-Cygni profile, even at high resolution, indicates that we are only probing the CSM, which displays only moderate expansion velocity, while the higher-velocity ejecta are hidden by this optically thick gas.

{As mentioned above, it is plausible that ILRTs are EC SNe arising from super-AGB stars, but there is no consensus on this topic yet. Due to this fundamental uncertainty the theoretical modelling of these object is still in a preliminary stage.
Different mechanisms have been suggested to power the ILRTs, including mass accretion on a main sequence star \cite{SokerMass2011}, a failed supernova due to fallback on a black hole \cite{Tsuna2020} and jets interacting with CSM \cite{Soker2021RAA....21...90S}. Some of these models envision ILRTs and LRNe arising from the same physical phenomenon. In any case, no model has been systematically used to describe ILRTs.
On the other hand, EC SN models are in a more advanced stage: thanks to simulations accounting for hydrodynamics and radiation transfer in the ejecta, recent studies are able to provide predictions for key observables of EC SNe, in particular, broad band light curves (e.g., \citep{Kozyreva2021MNRAS.503..797K}). So far, however, the results of such studies are more compatible with low-luminosity SNe IIP rather than with ILRTs. One key detail that may be lacking to explain ILRTs as EC SN, in the context of these theoretical models, is the presence of thick CSM surrounding the exploding star. Further studies both on the theoretical and the observational side are required to reach a better understanding of these transients.}

%%%%%%%%%%%%%%%%%%%%%%%%%%%%%%%%%%%%%%%%%%
\section{Luminous Red Novae}  

Luminous red novae (LRNe) are another important subclass of gap transients. In the past few years, we have observed about 20 {extragalactic events.}  
%which are typically brighter than the Galactic {\bf luminous} red novae \citep[e.g., V1309 Sco, V838 Mon; ][]{Goranskii2002AstL...28..691G, Munari2002A&A...389L..51M, Tylenda2005A&A...436.1009T, Mason2010A&A...516A.108M, Tylenda2011A&A...528A.114T}. 
The remarkable cases of this class are as follows: M31-RV, NGC4490-2011OT1, NGC3437-2011OT1, UGC12307-2013OT1, AT 2014ej, SNhunt248 %MDPI: we change footnote to Note format, please confirm.
\endnote{SNhunt248 is a LRN candidate whose nature is debated \citep{Kankare2015A&A...581L...4K}. }, M31-LRN2015, AT 2015dl/M101-2015OT1, AT 2017jfs, AT 2018bwo, AT 2018hso, AT 2019zhd, AT 2020hat, AT 2020kog, AT 2020hat, AT 2021biy, {AT\,2021afy and  AT\,2021blu}. \citep[][]{VanDyk2000PASP..112.1532V, Li2002PASP..114..403L, Boschi2004A&A...418..869B,Tully2009AJ....138..323T, Mason2010A&A...516A.108M,Schlafly2011ApJ...737..103S,Sorce2014MNRAS.444..527S, Adams2015MNRAS.452.2195A, Kankare2015A&A...581L...4K,Mauerhan2015MNRAS.447.1922M,Goranskij2016AstBu..71...82G,Smith2016MNRAS.458..950S,Blagorodnova2017ApJ...834..107B,Cai2019A&A...632L...6C,Pastorello2019A&A...625L...8P,Pastorello2019A&A...630A..75P,Stritzinger2020A&A...639A.104S,Pastorello2021A&A...646A.119P,Pastorello2021A&A...647A..93P,Blagorodnova2021A&A...653A.134B, Cai2022arXiv220700734C, Pastorello2022arXiv220802782P}. Most LRNe are located in spiral to irregular galaxies.

\begin{figure}[H] %H
\begin{adjustwidth}{-\extralength}{0cm}
\centering
\includegraphics[width=9.0cm]{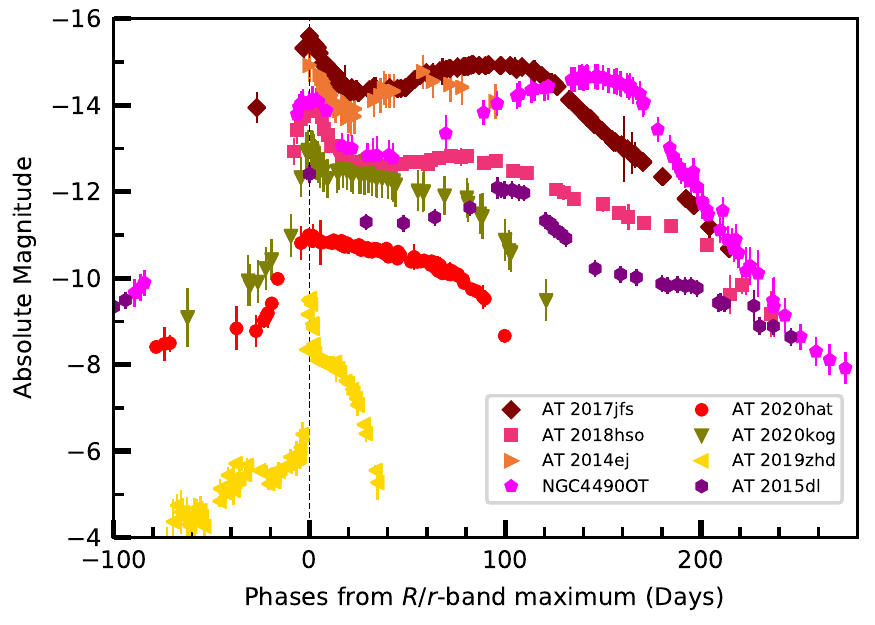}
\includegraphics[width=9.0cm]{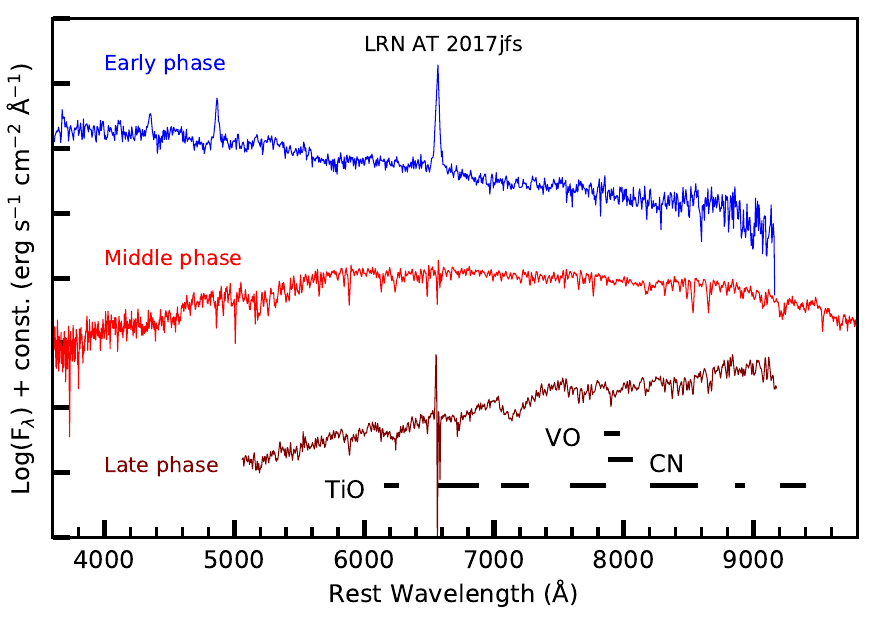}
\end{adjustwidth}
\caption{\textls[-45]{\textbf{Left} panel: $R/r$-band absolute light curves of LRNe. \textbf{Right} panel: spectral evolution of a remarkable LRN AT~2017jfs at early, middle, and late phases. Data from \citet{Pastorello2019A&A...625L...8P, Cai2019A&A...632L...6C,Smith2016MNRAS.458..950S, Blagorodnova2017ApJ...834..107B,Stritzinger2020A&A...639A.104S,Pastorello2021A&A...646A.119P,Pastorello2021A&A...647A..93P}.}
\label{LRNe}}
\end{figure}

{The Galactic luminous red nova V1309 Sco (e.g., \cite{Tylenda2011A&A...528A.114T,Pejcha2014ApJ...788...22P,Smith2016MNRAS.458..950S,Mason2010A&A...516A.108M}) was shown to be a short-period contact binary before its outburst, displaying a slow rise superposed on a periodic modulation (P$\sim$1.4 days) between 2002 and mid-2007 \cite{Pejcha2016MNRAS.455.4351P}. This period decreased with time as a consequence of the loss of orbital angular momentum \citep[][]{Tylenda2011A&A...528A.114T}.} Then, its luminosity started to decline following an exponential law {from late 2007 to early 2008, along with the disappearance of period variability.}  This is likely due to the orbital shrinking of the two stars which were therefore engulfed in a common envelope \citep{Tylenda2011A&A...528A.114T,Pejcha2016MNRAS.461.2527P}. Soon after this minimum, the light curve experiences a steep rise with a $\sim$3.5 mag brightening within $\sim$5 months {(from  March to late August 2008)}, which can be associated with the ejection of the common envelope. {The 7-year pre-outburst light curve of V1309 Sco observed by OGLE survey is well matched with the predictions for a contact binary.}
Unfortunately, extra-Galactic LRNe usually lack of deep photometry observations during the pre-outburst phases. So far, only two Galactic {luminous} red novae {(V1309 Sco (\citep{Tylenda2011A&A...528A.114T}, see their Figure~\ref{fig1}) and V838 Mon (\citep{Goranskij2004IBVS.5511....1G}, see their Figure \ref{fig1}))} have been {relatively} well-monitored during the pre-outburst phases. Finally, a huge brightening of $\sim$3.5 mag in less than two weeks was observed for V1309 Sco. Subsequently, these transients display a double-hump light curve typical of most LRNe (see the left panel of Figure~\ref{LRNe}).
The early short-duration peak, usually with a blue colour, is then followed by the second broad peak (or a plateau), characterised with a progressively redder colour. Afterwards, LRNe light curves usually experience a fast luminosity drop in all bands. Occasionally, a very late-time bump in the light curves can be observed for some LRNe (e.g., AT 2017jfs, AT 2021biy; \citep{Pastorello2019A&A...625L...8P, Cai2022arXiv220700734C}). 
Nonetheless, the theoretical mechanisms regulating the structured light curve of LRNe is still very poorly understood. In analogy to the plateau feature in Type IIP SNe \citep{Popov1993ApJ...414..712P}, hydrogen recombination might be the reason to stabilise the luminosity (e.g., \citep{Ivanova2013Sci...339..433I, Lipunov2017MNRAS.470.2339L}). 
However, recombination energy cannot give a major contribution to the first blue peak, as mentioned by \citet{MacLeod2017ApJ...835..282M}.
Instead, \citet{Metzger2017MNRAS.471.3200M} propose that the two-part structure light curve originates from the interaction between a fast ejected shell and a pre-existing equatorial wind. { Recently, \citet{Matsumoto2022arXiv220210478M} presented a new one-dimensional model of LRN light curves, which accounts for various effects of opacities, radiation, gas pressure, and the hydrogen recombination. This model can be applied to LRNe to infer the ejecta properties, such as ejecta mass, ejecta velocity, and launching radius.} High-dimension hydrodynamical simulations will be a powerful tool to solve these puzzles. 

During the pre-outburst phase, LRNe have never been observed in spectroscopy. Fortunately, some LRNe (e.g., NGC4490-2011OT1, AT~2017jfs, AT 2021biy) are well-monitored in optical spectroscopy during the major outburst phase in which they show similar spectral properties. Specifically, early-time (during the first peak) spectra show a blue continuum dominated by prominent H lines in emission, resembling those of other gap transients. During the second peak or plateau phase, the spectral continuum becomes progressively redder, similar to that of a K-type star. The weaker \Ha and a forest of narrow metal lines in absorption are observed in LRNe by that time. At very late phase, the spectra show an extremely red continuum resemble that of an M-type star. \Ha~becomes prominent again in pure emission, and the characterising molecular features (e.g., TiO, VO) emerge at such late phase. This dramatic spectral evolution behaviour (see the right panel of Figure~\ref{LRNe}) can be a diagnostic tool to discriminate LRNe from other gap transients, such as LBV outbursts and~ILRTs.

Several authors (e.g., \citep{Kochanek2014MNRAS.443.1319K, Pejcha2014ApJ...788...22P, Mauerhan2018MNRAS.473.3765M, Pastorello2019A&A...630A..75P, Pastorello2021A&A...647A..93P, Blagorodnova2021A&A...653A.134B, Pastorello2022arXiv220802782P}) looked for the possible correlations among the photometric and spectroscopic observables for LRNe. They obtained similar results: higher-luminosity events have longer evolution; hotter, larger, and higher-velocity photospheres; along with luminous \Ha~lines. In addition, the quiescent progenitors of bright events are generally more massive than dim ones. {(see their Equations (2) and (3), \mbox{\citet{Cai2022arXiv220700734C}}) obtained an empirical relation between the progenitor mass and the $V$-band absolute magnitudes at the second peak or plateau, which can be used for roughly inferring the progenitor mass of LRNe which lack detection of the progenitor in pre-explosion images.} Robust correlations in observational parameters can be provided with the expanding LRNe sample in future surveys.

\citet{Sana2012Sci...337..444S} provide an estimation that over 70\%~of massive stars interact with their companions and finally give rise to a binary merger with a ratio of about 30\%~of them. As presented by \citet{Kochanek2014MNRAS.443.1319K}, the stellar merger rate in the Galaxy is at the order of 0.5 yr %MDPI: should yr be changed to year?.
$^{-1}$ for events more luminous than $M_V=-3$ mag, while brighter events (\mbox{$M_V \leq -10$ mag}) have a much lower rate with $\sim$0.03 yr$^{-1}$.
Recently, \citet{Howitt2020MNRAS.492.3229H} estimated a Galactic rate of  an LRNe of about 0.2 yr$^{-1}$ and a volumetric LRNe rate of about 8 $\times$ 10$^{-4}$ Mpc$^{-3}$yr$^{-1}$ in the local Universe. Although the precise LRNe rate is still not available, we are confident that the intrinsic rate of LRNe is strongly dependent on their luminosity function and the mass of the progenitor binary system (e.g., \citep{Sana2012Sci...337..444S,Kochanek2014MNRAS.443.1319K,Howitt2020MNRAS.492.3229H,Pastorello2021A&A...647A..93P}). 

{There are several scenarios to interpret the LRN phenomenon: the thermonuclear runaway occurred on a low-mass degenerate dwarf was proposed to explain M31 RV by \citet{Iben1992ApJ...389..369I} or post-AGB star helium flash scenario (see e.g., \citep{Munari2002A&A...389L..51M, Kimeswenger2002MNRAS.336L..43K}). \citet{Pastorello2019A&A...630A..75P} also summarised several scenarios for the LRNe phenomenon, in which the major instabilities of single massive stars and the common envelope ejection followed by a stellar merger scenario are two popular ones. The watershed event is the Galactic LRN V1309 Sco which provided the strongest evidence that LRNe originate from the mergers of non-degenerate binaries (e.g., \citep{Mason2010A&A...516A.108M, Tylenda2011A&A...528A.114T}). For example, V838 Mon has been proposed to be a merger of two main sequence stars by \citet[][]{Soker2003ApJ...582L.105S}. On the other hand, the post-outburst outcome is still debated, as it can either result in a core-collapse SN event, or the system could survive as close binary stars (e.g., see \citep{Blagorodnova2017ApJ...834..107B, Pastorello2019A&A...630A..75P}). Very late-time photometry, in particular the IR observations, and spectroscopy will be needed to unveil the mystery. }

\section{Unknown Transients}

Modern, wide-field, time-domain surveys launched an upheaval in the astronomical field of studying transients, such as the discovery of kilonovae, {the electromagnetic counterpart of the merger of binary neutron stars} (e.g., \cite{Arcavi2017Natur.551...64A, Smartt2017Natur.551...75S, Wang2017ApJ...851L..18W, Xiao2017ApJ...850L..41X}). These surveys allow us to explore new regimes in the transients phase space represented in Figure~\ref{fig1}, which may also challenge the existing models for stellar evolution and explosions. Along with SN impostors, ILRTs, and LRNe, we are prepared to identify any possible transients lying in this luminosity gap, such as very faint and failed SNe (e.g., \cite{Lovegrove2013ApJ...769..109L}). 

%%%%%%%%%%%%%%%%%%%%%%%%%%%%%%%%%%%%%%%%%%
\section{Conclusion and Outlook}

Although significant progress has been made in both the follow-up and theoretical modelling for gap transients, our study is still in its early stages. The major obstacles are the small sample size and the lack of wavelength coverage in domains outside the optical and NIR ones. 
However, both issues will be soon effectively tackled: with the inauguration of {the Vera C. Rubin Observatory} (e.g., \citep{Bianco2022ApJS..258....1B, Breivik2022arXiv220802781B, Hambleton2022arXiv220804499H}), the sample size problem will be immediately solved, thanks to the large amount of data that this survey will provide. On the other hand, with the launch of the James Webb Space Telescope \citep{JWST2006}, it will be possible to investigate in the infrared domain in an unprecedented fashion, unveiling the secrets of the dust that is such a key component for many of these gap transients. 
In addition, comparing archival pre-outburst images at the transient site with ones taken several years after will discriminate whether the gap transients are non-terminal events or final explosions. With the advent of new facilities, high-cadence observations can help to cover their entire evolution. 
In particular, deeper photometry and intermediate-resolution spectroscopy performed on 2--4 m mid-size class telescopes are key for achieving such a goal. Late-time spectra obtained from 10 m class telescopes are another powerful tool to probe the physics of these events. On the other hand, the development of theoretical models will give helpful elements to understand the physical mechanisms and provide observational predictions for known and unknown transients in the gap. All of these efforts will allow us to better comprehend the nature of gap transients.

%\subsection{Figures, Tables and Schemes}

%All figures and tables should be cited in the main text as Figure~\ref{fig1}, Table~\ref{tab1}, Table~\ref{tab2}, etc.

%%\begin{table}[H] 
%%\caption{This is a table caption. Tables should be placed in the main text near to the first time they are~cited.\label{tab1}}
%%\begin{tabularx}{\textwidth}{CCC}
%%\toprule
%%\textbf{Title 1}	& \textbf{Title 2}	& \textbf{Title 3}\\
%%\midrule
%%Entry 1		& Data			& Data\\
%%Entry 2		& Data			& Data\\
%%\bottomrule
%%\end{tabularx}
%%\end{table}
%%\unskip

%%\begin{table}[H]
%%\caption{This is a wide table.\label{tab2}}
%%	\begin{adjustwidth}{-\extralength}{0cm}
%%		\begin{tabularx}{\fulllength}{CCCC}
%%			\toprule
%%			\textbf{Title 1}	& \textbf{Title 2}	& \textbf{Title 3}     & \textbf{Title 4}\\
%%			\midrule
%%			Entry 1		& Data			& Data			& Data\\
%%			Entry 2		& Data			& Data			& Data \textsuperscript{1}\\
%%			\bottomrule
%%		\end{tabularx}
%%	\end{adjustwidth}
%%	\noindent{\footnotesize{\textsuperscript{1} This is a table footnote.}}
%%\end{table}

\vspace{6pt} 

%%%%%%%%%%%%%%%%%%%%%%%%%%%%%%%%%%%%%%%%%%
%% optional
%\supplementary{The following supporting information can be downloaded at:  \linksupplementary{s1}, Figure S1: title; Table S1: title; Video S1: title.}


%%%%%%%%%%%%%%%%%%%%%%%%%%%%%%%%%%%%%%%%%%
\authorcontributions{Y.C. is the first author who is responsible for this paper; A.R. contributed to the Section of Supernova Impostors; G.V. contributed to the Section of Intermediate Luminosity Red Transients; X.W. supervised/supported Y.C. for this paper.
All authors have read and agreed to the published version of the manuscript.}

\funding{Y.C. is funded by the China Postdoctoral Science Foundation (grant no. 2021M691821). This work is supported by the National Natural Science Foundation of China (NSFC grants 12033003, 11633002), the Scholar Program of Beijing Academy of Science and Technology (DZ:BS202002), and the Tencent Xplorer Prize. A.R. acknowledges support from ANID BECAS/DOCTORADO NACIONAL 21202412.}

\institutionalreview{Not applicable}
%In this section, you should add the Institutional Review Board Statement and approval number, if relevant to your study. You might choose to exclude this statement if the study did not require ethical approval. Please note that the Editorial Office might ask you for further information. Please add “The study was conducted in accordance with the Declaration of Helsinki, and approved by the Institutional Review Board (or Ethics Committee) of NAME OF INSTITUTE (protocol code XXX and date of approval).” for studies involving humans. OR “The animal study protocol was approved by the Institutional Review Board (or Ethics Committee) of NAME OF INSTITUTE (protocol code XXX and date of approval).” for studies involving animals. OR “Ethical review and approval were waived for this study due to REASON (please provide a detailed justification).” OR “Not applicable” for studies not involving humans or animals.}

\informedconsent{Not applicable}
%Any research article describing a study involving humans should contain this statement. Please add ``Informed consent was obtained from all subjects involved in the study.'' OR ``Patient consent was waived due to REASON (please provide a detailed justification).'' OR ``Not applicable'' for studies not involving humans. You might also choose to exclude this statement if the study did not involve humans.

%Written informed consent for publication must be obtained from participating patients who can be identified (including by the patients themselves). Please state ``Written informed consent has been obtained from the patient(s) to publish this paper'' if applicable.}

\dataavailability{Some data are obtained from the Weizmann Interactive Supernova Data Repository (WISeREP) at \url{https://wiserep.weizmann.ac.il/}. %MDPI: Please add the access date (Format: Date Month Year). e.g., (accessed on 1 January 2020).
 Some data used in this work are available from the published literature. }
%{In this section, please provide details regarding where data supporting reported results can be found, including links to publicly archived datasets analyzed or generated during the study. Please refer to suggested Data Availability Statements in section ``MDPI Research Data Policies'' at \url{https://www.mdpi.com/ethics}. If the study did not report any data, you might add ``Not applicable'' here.} 

\acknowledgments{{The authors greatly thank the referees {and the academic editor} for their helpful comments and suggestions, that allowed us to improve the quality of the paper.} Y.C. thanks the instructive comments from the former PhD supervisor A. Pastorello.  } 

\conflictsofinterest{The authors declare no conflict of interest. The funders had no role in the design of the study; in the collection, analyses, or interpretation of data; in the writing of the manuscript, or in the decision to publish the~results.}
%Declare conflicts of interest or state ``The authors declare no conflict of interest.'' Authors must identify and declare any personal circumstances or interest that may be perceived as inappropriately influencing the representation or interpretation of reported research results. Any role of the funders in the design of the study; in the collection, analyses or interpretation of data; in the writing of the manuscript, or in the decision to publish the results must be declared in this section. If there is no role, please state ``The funders had no role in the design of the study; in the collection, analyses, or interpretation of data; in the writing of the manuscript, or in the decision to publish the~results''.} 

%%%%%%%%%%%%%%%%%%%%%%%%%%%%%%%%%%%%%%%%%%
%% Optional
%\sampleavailability{Samples of the compounds ... are available from the authors.}

%% Only for journal Encyclopedia
%\entrylink{The Link to this entry published on the encyclopedia platform.}

\abbreviations{Abbreviations}{
The following abbreviations are used in this manuscript:\\

\noindent 
\begin{tabular}{@{}ll}
MDPI & Multidisciplinary Digital Publishing Institute\\
%DOAJ & Directory of open access journals\\
SNe  & Supernovae\\
{SL SNe} & Superluminous SNe  \\
CC SNe & Core-collapse supernovae  \\
EC SNe & Electron Capture Supernovae \\
TDE   & Tidal Disruption Event \\
LBVs  & Luminous Blue Variables \\
ILRTs & Intermediate-luminosity red transients \\
LRNe & Luminous red novae \\
AGB  & Asymptotic giant branch \\
FWHM & Full-width at half-maximum \\
CSM  & Circumstellar material \\
HST  & Hubble Space Telescope \\
NIR  & Near Infrared \\
ZTF  & Zwicky Transient Facility  \\
ATLAS & Asteroid Terrestrial-impact Last Alert System \\
Pan-STARRS & Panoramic Survey Telescope and Rapid Response System \\
{OGLE} & {Optical Gravitational Lensing Experiment} \\
%TLA & Three letter acronym\\
%LD & Linear dichroism
\end{tabular}
}

%%%%%%%%%%%%%%%%%%%%%%%%%%%%%%%%%%%%%%%%%%
%% Optional
%%\appendixtitles{no} % Leave argument "no" if all appendix headings stay EMPTY (then no dot is printed after "Appendix A"). If the appendix sections contain a heading then change the argument to "yes".
%%\appendixstart
%%\appendix
%%\section[\appendixname~\thesection]{}
%%\subsection[\appendixname~\thesubsection]{}
%%The appendix is an optional section that can contain details and data supplemental to the main text---for example, explanations of experimental details that would disrupt the flow of the main text but nonetheless remain crucial to understanding and reproducing the research shown; figures of replicates for experiments of which representative data are shown in the main text can be added here if brief, or as Supplementary Data. Mathematical proofs of results not central to the paper can be added as an appendix.

%%\begin{table}[H] 
%%\caption{This is a table caption.\label{tab5}}
%%\begin{tabularx}{\textwidth}{CCC}
%%\toprule
%%\textbf{Title 1}	& \textbf{Title 2}	& \textbf{Title 3}\\
%%\midrule
%%Entry 1		& Data			& Data\\
%%Entry 2		& Data			& Data\\
%%\bottomrule
%%\end{tabularx}
%%\end{table}

%%\section[\appendixname~\thesection]{}
%%All appendix sections must be cited in the main text. In the appendices, Figures, Tables, etc. should be labeled, starting with ``A''---e.g., Figure A1, Figure A2, etc.

%%%%%%%%%%%%%%%%%%%%%%%%%%%%%%%%%%%%%%%%%%
\begin{adjustwidth}{-\extralength}{0cm}
\printendnotes[custom]

\reftitle{References}

\end{adjustwidth}
\end{document}